\documentclass[conference]{IEEEtran}
\usepackage{cite}
\usepackage{amsmath,amssymb,amsfonts}
\usepackage{graphicx}
\usepackage{textcomp}
\usepackage{xcolor}
\usepackage{colortbl}
\definecolor{heatblue}{RGB}{40,90,180}
\newcommand{\cell}[2]{%
	\edef\tmp{\noexpand\cellcolor{heatblue!\the\numexpr #1*100/100\relax}}%
	\tmp{#2}}
\newcommand{\cellb}[2]{%
	\edef\tmp{\noexpand\cellcolor{heatblue!\the\numexpr #1*100/100\relax}}%
	\tmp\textcolor{white}{\textbf{#2}}}
\usepackage{enumitem}
\usepackage{xurl}
\usepackage[hidelinks]{hyperref}
\usepackage{booktabs}
\usepackage{float}
\usepackage{placeins}
\usepackage{multirow}
\usepackage{array}
\usepackage{tabularx}
\usepackage{makecell}
\usepackage[activate={true,nocompatibility},final,tracking=true,factor=1100,stretch=10,shrink=10]{microtype}

\begin{document}
	
    \title{Forensic Schema for Psychological Manipulation in Cyber Fraud: LLM-Driven Victim Reports Analysis}
	
    \author{
    \IEEEauthorblockN{Zikai Alex Wen\textsuperscript{\textdagger},
    Corrazon Ogot\textsuperscript{\textdagger},
    Juan Li\textsuperscript{\textdaggerdbl},
    Yan Bai\textsuperscript{\textdagger}}
    \IEEEauthorblockA{
    \begin{tabular}{@{}c@{\hspace{2em}}c@{}}
    \begin{tabular}[t]{c}
    \textsuperscript{\textdagger}\textit{School of Engineering and Technology} \\
    \textit{University of Washington}\\
    Tacoma, United States \\
    \{zkwen, cogot12, yanb\}@uw.edu
    \end{tabular}
    &
    \begin{tabular}[t]{c}
    \textsuperscript{\textdaggerdbl}\textit{Department of Computer Science} \\
    \textit{North Dakota State University}\\
    Fargo, United States \\
    j.li@ndsu.edu
    \end{tabular}
    \end{tabular}
    }
    }
	
	\maketitle
	\begin{abstract}
		Existing cybercrime classification schemas capture contact metadata and financial transactions but omit the psychological manipulation techniques perpetrators employ. We present a forensic schema (four categories, 35 questions) adding 11 manipulation indicators and cryptocurrency evidence fields to established forensic foundations. Applied to 10{,}994 victim reports via large language model (LLM)-driven annotation and validated against two human annotators (mean LLM-human $\kappa = 0.69$, matching inter-annotator $\kappa = 0.68$), the schema revealed a statistically distinct manipulation profile for each major fraud type (Cram\'{e}r's~$V$ up to $0.790$). A rationale-based evidence audit nonetheless exposed a \emph{forensic detail gap}: detection of manipulation techniques was reliable, but victim narratives varied widely in the actionable detail supporting each Yes answer, and blockchain-specific identifiers were nearly absent. These findings point to AI-assisted victim intake with schema-informed follow-up questions as the most direct way to close the gap. The tiered annotation strategy also provides a reusable template for LLM-based extraction from other forensic text domains.
	\end{abstract}
	
	
	%% ============================================================
	\section{Introduction}
	%% ============================================================
	
	Cyber fraud has evolved from opportunistic individual schemes into industrialized operations~\cite{franceschini2023Compound}. Organized syndicates such as the Prince (Taizi) Group operate sprawling scam compounds across countries, training trafficked workers with shared scripts to execute psychologically sophisticated fraud at scale~\cite{Office}. The industrialized nature of these operations suggests that they produce recurring behavioral signatures, also known as \textit{modus operandi patterns}. If these patterns can be systematically captured, investigators could link seemingly independent complaints to a single syndicate, an approach known as \textit{cross-case linkage}~\cite{woodhams2007Case}.
	
	Existing cybercrime classification schemas have not kept pace with this evolution. Current frameworks~\cite{donalds2019Cybercrime,tsakalidis2019Systematic} capture contact metadata and financial transactions but do not categorize the psychological manipulation techniques perpetrators employ, leaving modus operandi elements buried in narratives.
	
	Meanwhile, the persuasion and natural language processing community has produced taxonomies for exactly these manipulation tactics. For example, Cialdini's six principles of persuasion~\cite{cialdini1993Influencea} describe how authority impersonation and scarcity pressure exploit cognitive biases, and PsyScam~\cite{ma2025PsyScam} benchmarked nine such techniques in real scam reports. However, these taxonomies remain disconnected from forensic applications. They identify \textit{what} perpetrators do but not \textit{how} that maps to evidence dimensions, investigative workflows, or prosecution needs. Since no existing framework bridges these perspectives, we ask the following two research questions:
	
	\begin{itemize}[leftmargin=*]
		\item \textbf{RQ1 (Forensic Associations):} What forensic associations exist between psychological manipulation profiles and fraud typology in victim reports?
		\item \textbf{RQ2 (Forensic Information Bottlenecks):} What information bottlenecks in victim reports limit the forensic utility of LLM-extracted manipulation indicators?
	\end{itemize}
	
	To answer these questions, we designed a forensic schema of 4 categories and 35 questions. Categories 1--2 establish the forensic foundation drawn from existing schemas~\cite{donalds2019Cybercrime,tsakalidis2019Systematic}. Category~3 (11 questions) breaks down the modus operandi into specific psychological manipulation tactics, each grounded in persuasion theory~\cite{cialdini1993Influencea,kahneman1979Prospect} and formulated as an answerable forensic question. Category~4 (5 questions) addresses a second gap in existing frameworks. As organized syndicates increasingly route proceeds through cryptocurrency~\cite{lim2025Modus}, investigators need blockchain-specific evidence such as wallet addresses, transaction hashes, and exchange identifiers that current schemas cannot structure. Category~4 structures these evidence elements as answerable questions, modeled on FBI IC3 intake fields~\cite{Cryptocurrency}.
	
	We applied the schema to 10{,}994 victim reports using a hybrid annotation methodology driven by a large language model (LLM), where each question was assigned to the extraction method matching the reasoning it required (Section~\ref{sec:methodology}). We validated annotation quality through structured-field cross-validation and human evaluation on a stratified subset.
	
	In summary, our work makes three contributions:
	\begin{enumerate}[leftmargin=*]
		\item A forensic schema bridging persuasion theory and forensic investigation across four evidence categories, with psychological manipulation profiling (Category~3) as the primary novel dimension.
		\item Empirical evidence from 10{,}994 victim reports showing that manipulation profiles distinguish fraud types (RQ1), but that victim narratives vary widely in actionable forensic detail (RQ2). This \emph{forensic detail gap} has three implications: (a)~the profiles constitute a behavioral feature space for automated fraud-type triage (Section~\ref{sec:triage}); (b)~schema-informed follow-up questions can dynamically elicit the missing detail during victim intake (Section~\ref{sec:intake}); and (c)~blockchain-specific identifiers are nearly absent, representing a critical bottleneck for cryptocurrency fraud investigation (Section~\ref{sec:intake}).
		\item A tiered annotation methodology that assigns each question to the extraction method matching its reasoning demands, providing a reusable template for LLM-based structured extraction from other forensic and security text domains (Section~\ref{sec:characterization}).
	\end{enumerate}
	
	%% ============================================================
	\section{Related Work}
	%% ============================================================
	We organize prior work along three dimensions: forensic reporting standards that structure cybercrime data, psychological theories that explain how perpetrators manipulate victims, and LLM-based methods that enable scalable text annotation. We then identify the gap our schema addresses.
	
	\subsection{Forensic Reporting Standards}
	
	Cybercrime classification schemas~\cite{donalds2019Cybercrime,tsakalidis2019Systematic}, standardized exchange formats such as CybOX~\cite{casey2015Leveraging}, and ontology-driven approaches~\cite{sikos2021AI,harichandran2016CuFA} have advanced incident standardization but focus on technical artifact representation, not psychological manipulation tactics. The need for richer standardization is reinforced by work on structured fraud report data~\cite{girocorreia2022Making} and inconsistencies in police cybercrime measurement~\cite{kwon2024Measuring}.
	
	On the investigative side, case linkage analysis identifies serial offenders through behavioral patterns~\cite{woodhams2007Case}, and digital forensic frameworks have evolved to support scalable evidence processing~\cite{vanbeek2020Digital} and fintech investigations~\cite{nikkel2020Fintech}. However, these methodologies have not been applied to cyber fraud, where behavioral patterns manifest as manipulation rather than physical traces.
	
	Two additional gaps motivate our work. First, victims often under-report~\cite{fonseca2022Online}, so reports may omit tactics that were actually used. Second, cryptocurrency fraud introduces evidential challenges such as wallet tracing and exchange attribution that existing frameworks do not address~\cite{frowis2020Safeguarding}. Our schema addresses both: Category~3 captures the manipulation patterns that behavioral case linkage requires, and Category~4 structures the blockchain evidence that existing frameworks lack.
	
	\subsection{Psychological Manipulation in Fraud}
	Cialdini's six principles of persuasion (i.e., reciprocity, commitment and consistency, social proof, authority, liking, and scarcity)~\cite{cialdini1993Influencea} provided a foundational framework for understanding how perpetrators exploit cognitive biases. At a structural level, routine activity theory explains that fraud opportunities arise when a motivated offender encounters a suitable target in the absence of capable guardianship, a condition that online environments readily create~\cite{pratt2010Routine}. In phishing research, authority and scarcity were found to be the most frequently exploited principles~\cite{khadka2023Survey}. In romance scams, liking and commitment exploitation were shown to follow identifiable temporal scripts~\cite{schokkenbroek2025Love,coluccia2020Online}, with recent work tracing the evolution of romance fraud into ``pig butchering'' schemes~\cite{cross2024Romance, lim2025Modus}.
	
	PsyScam~\cite{ma2025PsyScam} is the most directly relevant prior work. Ma et al.\ benchmarked nine psychological techniques in real-world scams grounded in Cialdini's principles, prospect theory~\cite{kahneman1979Prospect}, and the elaboration likelihood model~\cite{petty1986Elaboration}. However, PsyScam was designed for technique detection, not forensic investigation. It classifies \textit{what} tactic was used but not \textit{why} that matters for an investigator, and acknowledges taxonomic gaps such as enforced isolation, where the perpetrator instructs the victim to cut off outside advisors~\cite{ma2025PsyScam}. Surveys of computational persuasion~\cite{bozdag2026Musta} and LLM-driven persuasion~\cite{liu2025LLMa} similarly lack forensic grounding. Therefore, our Category~3 schema design addresses these limitations.
	
	\subsection{LLMs in Digital Forensics and Fraud Analysis}
	
	Large language models have been applied to a growing range of digital forensic tasks, including evidence triage and report generation~\cite{dunsin2024Comprehensive,wickramasekara2025Exploring}. Most relevant to our approach, Relins et al.~\cite{relins2025Using} applied instruction-tuned LLMs to annotate unstructured police incident narratives with domain-specific vulnerability indicators, demonstrating that LLMs can extract structured forensic categories from free-text law enforcement reports. Their work validated the core premise of our methodology, that LLMs can perform schema-guided annotation of investigative text, though they targeted physical-world vulnerability indicators rather than psychological manipulation in cyber fraud. A recent review of computational text analysis of police data~\cite{lukmanjaya2026Computational} confirmed that fraud-specific applications of these methods remain largely unaddressed.
	
	More broadly, Gilardi et al.~\cite{gilardi2023ChatGPT} showed that LLM annotations can match crowd-worker quality, and Savelka et al.~\cite{savelka2023Unreasonable} demonstrated strong zero-shot performance on legal text annotation, supporting LLM use in forensic pipelines.
	
	\subsection{Gap Analysis}
	
	Table~\ref{tab:gap} summarizes our positioning. PsyScam provides the psychological taxonomy; cybercrime classification schemas and forensic exchange formats provide the forensic grounding. No prior work has integrated manipulation profiling with forensic grounding and validation on real victim reports.
	
	\begin{table}[t]
		\centering
		\caption{Positioning relative to existing work.}
		\label{tab:gap}
		\footnotesize
		\begin{tabular}{@{}p{3.5cm}ccc@{}}
			\toprule
			& \textbf{Psych.} & \textbf{Forensic} & \textbf{Victim} \\
			& \textbf{Taxonomy} & \textbf{Grounding} & \textbf{Reports} \\
			\midrule
			Cialdini~\cite{cialdini1993Influencea} & \checkmark & & \\
			PsyScam~\cite{ma2025PsyScam} & \checkmark & & \checkmark \\
			Donalds \& Osei-Bryson~\cite{donalds2019Cybercrime} & & \checkmark & \\
			Tsakalidis \& Vergidis~\cite{tsakalidis2019Systematic} & & \checkmark & \\
			\textbf{Our Work} & \checkmark & \checkmark & \checkmark \\
			\bottomrule
		\end{tabular}
	\end{table}
	
	%% ============================================================
	\section{Schema Design}
	\label{sec:schema}
	%% ============================================================
	
	The schema comprises 4 categories and 35 questions, all using a ternary Yes/No/Unknown (Y/N/U) answer format. Table~\ref{tab:cat_summary} summarizes all four categories. The categories progress from established forensic foundations to our new contributions:
	
	\begin{itemize}[leftmargin=*]
		\item \textbf{Category~1: Standard Forensic Metadata} (16 questions). Contact and subject identification fields, along with key financial metadata (i.e., transaction dates, amounts, total loss, payment method, and routing instructions), drawn from cybercrime classification schemas~\cite{donalds2019Cybercrime,tsakalidis2019Systematic}.
		\item \textbf{Category~2: Traditional Financial Flow} (3 questions). Traditional banking involvement and recipient bank identification, included for schema completeness~\cite{donalds2019Cybercrime,tsakalidis2019Systematic}.
		\item \textbf{Category~3: Psychological Manipulation Indicators} (11 questions). The schema's primary contribution: a structured modus operandi taxonomy grounded in persuasion theory~\cite{cialdini1993Influencea,kahneman1979Prospect} and translated into answerable forensic questions.
		\item \textbf{Category~4: Cryptocurrency Evidence} (5 questions). As organized fraud syndicates increasingly route proceeds through digital assets~\cite{lim2025Modus}, investigators need blockchain-specific evidence (wallet addresses, transaction hashes, cryptocurrency type) that existing classification schemas do not capture. The FBI IC3 has recognized this need by introducing cryptocurrency-specific intake fields~\cite{Cryptocurrency}; we model Category~4 on these fields to assess whether victim reports provide actionable blockchain detail.
	\end{itemize}
	
	We detail the design of Categories~3--4 in the following subsections and list the Category~1--2 questions in Appendix~\ref{app:cat12}.

	\begin{table}[t]
		\centering
		\caption{Schema summary by category.}
		\label{tab:cat_summary}
		\footnotesize
		\begin{tabular}{@{}clcc@{}}
			\toprule
			\textbf{Cat.} & \textbf{Description} & \textbf{Qs} & \textbf{Source} \\
			\midrule
			1 & Standard Forensic Metadata & 16 & \cite{donalds2019Cybercrime,tsakalidis2019Systematic} \\
			2 & Traditional Financial Flow & 3 & \cite{donalds2019Cybercrime,tsakalidis2019Systematic} \\
			3 & Psych.\ Manipulation Indicators & 11 & Ours \\
			4 & Cryptocurrency Evidence & 5 & Ours \\
			\midrule
			\multicolumn{2}{@{}l}{\textbf{Total}} & \textbf{35} & \\
			\bottomrule
		\end{tabular}
	\end{table}
	
	\subsection{Category~3: Psychological Manipulation Indicators}
	\label{sec:cat3}
	
	Current forensic intake instruments, including the FBI IC3 complaint form~\cite{FAQ,Cryptocurrency}, treat modus operandi as a single narrative field. A case where the perpetrator impersonated a government official and one where the perpetrator built a romantic relationship over months are recorded the same way. These distinctions matter: impersonating an official supports aggravated charges and agency referral, while romantic grooming indicates a different fraud typology, victim vulnerability profile, and temporal pattern.
	
	To achieve forensic granularity within the modus operandi field, we developed 11 indicators inspired by PsyScam's psychological technique taxonomy~\cite{ma2025PsyScam} and grounded in Cialdini's principles of persuasion~\cite{cialdini1993Influencea} and prospect theory~\cite{kahneman1979Prospect}. Prior work validated that these psychological constructs manifest in real-world fraud~\cite{khadka2023Survey,schokkenbroek2025Love}. Our contribution is to translate them into answerable forensic questions through the following design steps:
	
	\begin{itemize}[leftmargin=*]
		\item \textbf{Forensic Justification:} Each psychological technique is paired with an explicit forensic rationale, specifying why the technique matters for investigation or prosecution and what evidence dimension it maps to.
		\item \textbf{Forensic Extension:} Beyond the nine core persuasion-based indicators, we added two indicators addressing gaps acknowledged in the persuasion literature~\cite{ma2025PsyScam}: \textit{Communication Migration} (i.e., capturing instructions to move to private channels) and \textit{Post-Payment Behavioral Shift} (i.e., capturing contact severance and communication pattern changes after payment).
	\end{itemize}
	
	\begin{table}[t]
		\centering
		\caption{Category~3: Forensic psychological manipulation indicators (Q3.1--3.11).}
		\label{tab:cat3}
		\footnotesize
		\begin{tabular}{@{}cp{6.2cm}@{}}
			\toprule
			\textbf{Q\#} & \textbf{Operationalized Question} \\
			\midrule
			\multicolumn{2}{@{}l}{\textit{Persuasion-theory-based indicators:}} \\
			3.1 & Did the perpetrator claim to represent a government, law enforcement, or corporate entity? \\
			3.2 & Did the perpetrator induce fear through threats of legal action, arrest, or penalties? \\
			3.3 & Did the perpetrator create time pressure through deadlines or limited availability? \\
			3.4 & Did the perpetrator build rapport through romantic interest, flattery, or emotional connection? \\
			3.5 & Did the perpetrator impersonate a real entity or construct false credibility using fabricated artifacts (e.g., fake documents, scam websites)? \\
			3.6 & Did the perpetrator provide initial benefits to create obligation? \\
			3.7 & Did the perpetrator exploit commitment through incremental escalation or post-payment fee demands (e.g., taxes, unfreezing fees, withdrawal fees)? \\
			3.8 & Did the perpetrator use social proof to influence the victim (e.g., others' success stories)? \\
			3.9 & Did the perpetrator promise unrealistic returns, winnings, or financial rewards? \\
			\midrule
			\multicolumn{2}{@{}l}{\textit{Forensic extensions:}} \\
			3.10 & Did the perpetrator instruct the victim to migrate communication to private channels? \\
			3.11 & Did the perpetrator become unreachable or sever contact after receiving payment? \\
			\bottomrule
		\end{tabular}
	\end{table}
	
	Table~\ref{tab:cat3} presents the resulting 11 indicators. We clarify what each captures in practice in the following paragraphs.
	
	\textit{Authority} (Q3.1) marks cases where the perpetrator claims to represent a specific organization, such as a bank, a government agency, or a law enforcement body. \textit{Fear} (Q3.2) captures explicit threats of serious consequences: arrest, lawsuits, or account seizure. These two often co-occur, but Authority can appear without Fear (e.g., a perpetrator posing as a tech-support agent offering help) and Fear without Authority (e.g., a sextortion threat from an anonymous sender). \textit{Urgency} (Q3.3) captures imposed deadlines or demands for immediate action, such as ``your account will be locked in 24 hours.''
	
	\textit{Liking} (Q3.4) captures deliberate rapport-building through romantic interest, flattery, or emotional bonding. \textit{Pretext} (Q3.5) marks the fabrication of artifacts designed to appear legitimate: counterfeit credentials, forged documents, or spoofed websites. Pretext and authority are related but distinct. Pretext is the deceptive artifact or evidence, while authority is the claim of a professional role. These two may overlap when a role claim invokes a named real organization. Investment or trading platform apps are excluded from Pretext because they constitute the fraud mechanism itself, not an artifact used to claim legitimacy.
	
	\textit{Reciprocity} (Q3.6) captures cases where the perpetrator personally gives something, such as a small payout, a gift, or intimate content, to create a sense of obligation. \textit{Consistency} (Q3.7) captures post-payment escalation: after an initial payment, the perpetrator demands additional money under new pretexts such as ``taxes,'' ``unfreezing fees,'' or ``withdrawal charges.'' These two indicators track different stages of financial exploitation: \textit{Reciprocity} is the hook, \textit{Consistency} is the ratchet. A subtle but important boundary separates \textit{Consistency} from \textit{Fear}: when a platform freezes a victim's funds and demands a fee to release them, it is \textit{Consistency}; when the freeze is accompanied by threats of legal action, it is \textit{Fear}.
	
	\textit{Social proof} (Q3.8) captures visible demonstrations that others have succeeded, such as screenshots of earnings or fabricated testimonials in group chats. \textit{Phantom riches} (Q3.9) captures promises of unusually high or guaranteed returns.
	
	The two forensic extensions address gaps acknowledged in persuasion taxonomies~\cite{ma2025PsyScam}. \textit{Communication migration} (Q3.10) captures instructions to move from a public or traceable channel to a private one, such as from a dating app to WhatsApp or a scammer-managed group chat, which constitutes evidence of consciousness of guilt. \textit{Post-payment behavioral shift} (Q3.11) marks the point where the perpetrator becomes unreachable or the platform goes offline after receiving payment, providing a temporal boundary that helps establish intent.
	
	\subsection{Category~4: Cryptocurrency Evidence}
	\label{sec:cat4}
	
	Category~4 structures the evidence elements investigators require as answerable questions, modeled on FBI IC3 intake fields~\cite{Cryptocurrency}. Table~\ref{tab:cat4} presents the 5 questions. \textit{Cryptocurrency involvement} (Q4.1) is a gating indicator that triggers the remaining four questions. \textit{Transaction hashes} (Q4.3) and \textit{recipient wallet addresses} (Q4.4) are critical because blockchain tracing techniques such as wallet clustering, exchange KYC matching, and cross-complaint address linking~\cite{frowis2020Safeguarding} require these on-chain identifiers. \textit{Crypto ATM involvement} (Q4.5) captures a physical dimension yielding surveillance footage and geolocation. 
	
	\begin{table}[t]
		\centering
		\caption{Category~4: Cryptocurrency evidence questions with forensic rationale.}
		\label{tab:cat4}
		\footnotesize
		\begin{tabular}{@{}cp{6.2cm}@{}}
			\toprule
			\textbf{Q\#} & \textbf{Operationalized Question} \\
			\midrule
			4.1 & Did the transaction involve cryptocurrency? \\
			4.2 & Is the type of cryptocurrency known? \\
			4.3 & Is the transaction hash available? \\
			4.4 & Is the recipient's wallet address available? \\
			4.5 & Was a crypto ATM or kiosk used for the transaction? \\
			\bottomrule
		\end{tabular}
	\end{table}

	%% ============================================================
	\section{Methodology}
	\label{sec:methodology}
	%% ============================================================
	
	This section describes how we applied the schema to real victim reports. We first introduce the dataset, then detail the hybrid annotation strategy that assigns each schema question to the appropriate extraction method, and finally describe the human evaluation and statistical tests used to answer RQs.
	
	\subsection{Dataset}
	
	The corpus comprises 10{,}994 cyber fraud victim reports\footnote{The dataset is available at \url{https://research.zkwen.site/scamschema}.} drawn from eight publicly available sources spanning multiple jurisdictions. Five are U.S.\ consumer protection and regulatory scam trackers: BBB~\cite{Search}, California DFPI~\cite{Crypto}, Washington State DFI~\cite{Washington}, Wisconsin DFI~\cite{DFI}, and DC DISB~\cite{DISB}. Three are academic datasets: the CCL23-Eval Task~6 telecom fraud corpus~\cite{sun2023Overview}, the crime script analysis dataset of Lwin Tun and Birks~\cite{lwintun2023Supporting}, and the PsyScam benchmark~\cite{ma2025PsyScam}. 
	
	Reports span 7 consolidated fraud categories derived from FBI Common Frauds and Scams Categories~\cite{Common}: Business and Investment Fraud (27.5\%), Spoofing and Phishing (25.2\%), Romance Scams (21.4\%), Consumer Fraud Schemes (11.4\%), Cryptocurrency Investment Fraud (7.3\%), Sextortion (4.1\%), and Job Scams (3.0\%).
	
	Each report is represented by 22 structured fields: a unique case identifier, report date, and fraud type label; a free-text victim narrative (i.e., description); contact metadata (i.e., platform, method, date, scammer name, phone, email, and website); and financial transaction details (i.e., date, amount, currency, recipient name, bank name, account number, and payment method). For cases involving cryptocurrency, additional fields record the wallet address, cryptocurrency type, cryptocurrency amount, and transaction hash.
	
	\subsection{Ethical Considerations}
	All data used in this study were collected from publicly available sources. The dataset contains no victim personally identifiable information.
	
	\subsection{Annotation}
	\label{sec:annotation}
	
	The 35 schema questions vary in the type of reasoning they require. Some can be resolved by direct field lookup, others require extracting information from free-text narratives, and a third group demands semantic interpretation beyond surface-level pattern matching. Accordingly, we employed a hybrid annotation strategy that assigned each question to the most appropriate method. The per-tier question assignments are detailed in Appendix~\ref{app:prompt}.
	
	\emph{Tier~1: Structured field lookup} (14 questions). We annotated deterministically those questions that mapped directly to consistently populated structured fields: if the field is non-empty, the answer is~Y with the field value; otherwise~N.
	
	\emph{Tier~2: Field lookup with narrative supplementation} (4 questions). Questions whose structured fields have incomplete coverage. We checked the field first; if empty, the LLM extracted the answer from the narrative.
	
	\emph{Tier~3: Narrative extraction} (6 questions). Questions with no corresponding structured field. The LLM extracted these directly from the narrative.
	
	\emph{Tier~4: Semantic interpretation} (11 questions). The entire Category~3 (Q3.1--Q3.11), which required inferential reasoning to recognize persuasion tactics such as authority impersonation, urgency framing, and escalation patterns.
	
	\subsection{LLM Setup} 
	
	Tier~2 through Tier~4 questions were annotated using the Claude Haiku 4.5 (\texttt{claude-haiku-4-5-20251001}) API. We followed the LLM-as-annotator methodology established by Gilardi et al.~\cite{gilardi2023ChatGPT}, who showed that LLM annotations can match or exceed crowd-worker quality on text classification tasks. In their experiments, the same model produced identical labels 97\% of the time when run twice on the same input. We configured the model with \texttt{temperature\,=\,0.2} to balance annotation consistency with output quality. Since Gilardi et al.\ used GPT-series models, we validated the approach for our model through self-agreement measurement and human evaluation (Section~\ref{sec:human_eval}).
	
	For each case, the LLM received a structured prompt containing the full schema with question definitions, the structured fields and victim narrative, and instructions to output a JSON array with Y/N/U labels and a short rationale for each indicator marked~Y. The per-indicator rationales serve as an audit trail for every annotation decision. A condensed version of the prompt\footnote{The full prompt is available at \url{https://research.zkwen.site/scamschema}.} is shown in Appendix~\ref{app:prompt}.
    
	We ensured annotation quality through two complementary checks: structured-field cross-validation against Tier~1 ground truth, and the human evaluation on a stratified sample described in the next subsection.
	
	\subsection{Human Evaluation}
	\label{sec:human_eval}
	
	To validate LLM annotation reliability for Tier~4 questions, two annotators (co-authors with training in digital forensics) independently labeled a stratified subset of 228 cases (2.1\% of the corpus). The subset maintained proportional representation across all seven fraud categories, with minority-category oversampling to ensure stable estimates. Cohen's $\kappa$ measures how often annotators agree beyond what chance would predict, on a scale from 0 (chance) to 1 (perfect). Table~\ref{tab:human_eval} reports the results. We found average human-human $\kappa$ of 0.68, which Landis and Koch~\cite{landis1977Measurement} classify as substantial agreement, and average LLM-human $\kappa$ of 0.69, on par with the human-human average. The two lowest human-human indicators, Liking (Q3.4, $\kappa = 0.48$) and Reciprocity (Q3.6, $\kappa = 0.49$), both had fewer than 30 positive cases, a regime where $\kappa$ is known to be unstable under skewed marginals. Overall, the LLM annotations were reliable enough to support the population-level statistical analyses in the following sections.

	\begin{table}[t]
		\centering
		\caption{Inter-annotator and LLM-human agreement for Category~3 indicators (Cohen's $\kappa$, $n = 228$ stratified cases).}
		\label{tab:human_eval}
		\footnotesize
		\begin{tabular}{@{}llcc@{}}
			\toprule
			\textbf{Q\#} & \textbf{Indicator} & \textbf{H-H $\kappa$} & \textbf{LLM-H $\kappa$} \\
			\midrule
			3.1 & Authority          & 0.88 & 0.84 \\
			3.2 & Fear               & 0.68 & 0.79 \\
			3.3 & Urgency            & 0.65 & 0.59 \\
			3.4 & Liking             & 0.48 & 0.57 \\
			3.5 & Pretext            & 0.72 & 0.80 \\
			3.6 & Reciprocity        & 0.49 & 0.42 \\
			3.7 & Consistency         & 0.76 & 0.74 \\
			3.8 & Social proof       & 0.67 & 0.62 \\
			3.9 & Phantom riches     & 0.75 & 0.73 \\
			3.10 & Comm.\ migration  & 0.67 & 0.69 \\
			3.11 & Post-payment shift & 0.72 & 0.83 \\
			\midrule
			& \textbf{Average}       & \textbf{0.68} & \textbf{0.69} \\
			\bottomrule
		\end{tabular}
	\end{table}
	
	\subsection{Statistical Analysis}
	
	We tested associations between Category~3 manipulation technique profiles and fraud typology using chi-square tests of independence between each indicator and the fraud typology, with Cram\'{e}r's $V$ for effect size. All tests used $\alpha = 0.001$ to control for multiple comparisons across the 11 indicators.
	
	We also conducted a rationale-based evidence audit to identify where victim narratives limit the forensic utility of LLM-extracted manipulation indicators. Our hybrid annotation pipeline produced, for every Y answer, a short rationale (15--60 words) quoting or closely paraphrasing the supporting evidence from the victim narrative. We used these rationales to assess evidence quality. These rationales revealed not only \emph{whether} the LLM detected a tactic but \emph{what evidence the narrative actually contained}.
	
	We assessed two dimensions. First, \emph{evidence thinness}: for each Category~3 indicator, we measured the proportion of Y rationales that contained forensically actionable details (specific platform names, monetary amounts, dates, or named entities) versus generic descriptions (e.g., ``perpetrator impersonated police'' with no further specifics). Second, \emph{indicator information ceiling}: we identified indicators where victim narratives provided too little signal for meaningful extraction, characterized by low Y-rates even in fraud types where the tactic is known to be common (e.g., Social proof in Cryptocurrency Investment Fraud).
	
	For Category~4, where the question is whether victims provided on-chain identifiers at all, we used a simpler coverage-based analysis. Q4.1 establishes whether cryptocurrency was involved. Q4.2--4.5 are then evaluated only among the crypto-involved cases (those where Q4.1 = Y, $N_{\text{crypto}} = 1{,}172$).
	
	%% ============================================================
	\section{Results}
	\label{sec:results}
	%% ============================================================
	
	\subsection{RQ1: Forensic Associations}
	
	We found that all 11 Category~3 indicators showed statistically significant associations with fraud typology (all $p < 0.001$). Cram\'{e}r's~$V$ measures how strongly an indicator's presence or absence is linked to the fraud type. Values above 0.5 indicate a large effect. The strongest associations were Pretext ($V = 0.685$), Post-payment shift ($V = 0.526$), Fear ($V = 0.520$), and Authority ($V = 0.504$). The weakest was Liking ($V = 0.230$), though still highly significant.
	Table~\ref{tab:rq1_fraud_type} presents all associations.
	
	\begin{table}[t]
		\centering
		\caption{Indicator-fraud type associations (chi-square tests, $N = 10{,}994$, all $p < 0.001$).}
		\label{tab:rq1_fraud_type}
		\footnotesize
		\begin{tabular}{@{}llcc@{}}
			\toprule
			\textbf{Q\#} & \textbf{Indicator} & \textbf{$\chi^2$} & \textbf{$V$} \\
			\midrule
			3.5 & Pretext & 5163.5 & 0.685 \\
			3.11 & Post-payment shift & 3040.5 & 0.526 \\
			3.2 & Fear & 2974.3 & 0.520 \\
			3.1 & Authority & 2796.2 & 0.504 \\
			3.8 & Social proof & 2214.6 & 0.449 \\
			3.7 & Consistency & 1612.0 & 0.383 \\
			3.10 & Comm.\ migration & 1163.0 & 0.325 \\
			3.9 & Phantom riches & 1006.1 & 0.303 \\
			3.6 & Reciprocity & 825.0 & 0.274 \\
			3.3 & Urgency & 760.8 & 0.263 \\
			3.4 & Liking & 579.6 & 0.230 \\
			\bottomrule
		\end{tabular}
	\end{table}
	
	The prevalence matrix (Table~\ref{tab:prevalence}) shows that fraud types differed not only in \emph{which} tactics are present but in \emph{how many}: Cryptocurrency Investment Fraud exhibited the highest tactic density (mean 3.6 per case), followed by Business and Investment Fraud (mean 2.5) and Sextortion (mean 2.2), while Consumer Fraud was leanest (mean 0.9).
	
	\subsubsection{Coercive profiles} Spoofing and Sextortion both use fear-based tactics, but their manipulation profiles are structurally distinct. In Spoofing cases, 73.5\% involve Authority claims (e.g., impersonating a bank or government agency) and 26.4\% involve Fear (e.g., threatening account closure), consistent with phishing research identifying authority as the most exploited principles~\cite{khadka2023Survey}. Sextortion relied on Fear far more heavily (81.3\% of cases) but almost never involved Authority claims (2.6\%); instead, it paired Fear with Consistency (48.9\% of cases, reflecting repeated payment demands) and Communication migration (46.7\%, reflecting channel isolation). In short, Spoofing is a single-interaction authority play; Sextortion is a sustained fear-and-escalation campaign.
	
	\subsubsection{Investment-based profiles} Cryptocurrency Investment Fraud used the widest range of manipulation tactics (mean 3.6 per case). In this fraud type, 85.1\% of cases involve fabricated credentials (Pretext), 58.5\% involve escalating fee demands (Consistency), 54.3\% involve the perpetrator disappearing after payment (Post-payment shift), and 42.7\% involve promises of high returns (Phantom riches). Business/Investment Fraud shared the escalation pattern (40.5\% Consistency) but relied instead on Post-payment shift (62.6\%) and third-party endorsements (Social proof, 33.2\%), with almost no platform fabrication (Pretext at only 2.7\%). These differences quantify how cryptocurrency-based pig-butchering schemes~\cite{cross2024Romance,lim2025Modus} depend on fabricated infrastructure and initial return payments in ways that traditional investment fraud does not.
	
	\subsubsection{Financial-gain profiles} Job Scams showed moderate escalation (Consistency in 42.8\% of cases) and channel isolation (Communication migration, 30.3\%), with unrealistic earnings promises (Phantom riches) in 12.2\%. Consumer Fraud presented the sparsest profile (mean 0.9 tactics per case), with Authority claims as the only indicator exceeding 25\%.
	
	\subsection{RQ2: Forensic Information Bottlenecks}
	
	RQ1 established that the LLM can reliably detect which manipulation tactics were used. RQ2 asks a different question: when the LLM labels a tactic as present, how much \emph{useful forensic detail} does the victim narrative actually contain?
	
	\begin{table*}[t]
		\centering
		\caption{Category~3 prevalence matrix: percentage of reports with indicator present (Y) by fraud type ($N = 10{,}994$). Darker shading indicates higher prevalence; cells $\geq$ 50\% are \textbf{bolded}.}
		\label{tab:prevalence}
		\footnotesize
		\setlength{\tabcolsep}{4pt}
		\renewcommand{\arraystretch}{1.4}
		\begin{tabular}{@{}cl*{7}{w{c}{0.085\textwidth}}@{}}
			\toprule
			& & \multicolumn{7}{c}{\textbf{Fraud Type (Yes \%)}} \\
			\cmidrule(l){3-9}
			\textbf{Q\#} & \textbf{Indicator} & \textbf{Spoofing} & \textbf{Romance} & \textbf{Consumer} & \textbf{Crypto Inv.} & \textbf{Sextortion} & \textbf{Bus. \& Inv.} & \textbf{Job Scams} \\
			\midrule
			3.1 & Authority      & \cellb{74}{73.5} & \cell{14}{14.4} & \cell{26}{25.6} & \cell{27}{27.0} & \cell{3}{2.6}  & \cell{24}{23.9} & \cell{14}{14.1} \\
			3.2 & Fear           & \cell{26}{26.4} & \cell{4}{3.9}  & \cell{7}{7.2}  & \cell{15}{14.5} & \cellb{81}{81.3} & \cell{1}{0.6}  & \cell{2}{1.8}  \\
			3.3 & Urgency        & \cell{21}{20.9} & \cell{13}{12.7} & \cell{8}{7.5}  & \cell{8}{7.8}  & \cell{3}{3.3}  & \cell{1}{0.6}  & \cell{1}{0.9}  \\
			3.4 & Liking         & \cell{3}{3.4}  & \cell{5}{4.9}  & \cell{3}{2.6}  & \cell{15}{14.6} & \cell{7}{7.3}  & \cell{18}{18.2} & \cell{6}{6.4}  \\
			3.5 & Pretext        & \cell{5}{5.4}  & \cell{4}{4.4}  & \cell{1}{1.1}  & \cellb{85}{85.1} & \cell{26}{26.4} & \cell{3}{2.7}  & \cell{18}{18.3} \\
			3.6 & Reciprocity    & \cell{5}{4.8}  & \cell{7}{7.1}  & \cell{9}{8.7}  & \cell{18}{17.9} & \cell{2}{2.4}  & \cell{27}{26.6} & \cell{19}{18.7} \\
			3.7 & Consistency    & \cell{7}{7.1}  & \cell{31}{30.5} & \cell{8}{7.6}  & \cellb{59}{58.5} & \cell{49}{48.9} & \cell{41}{40.5} & \cell{43}{42.8} \\
			3.8 & Social proof   & \cell{1}{0.9}  & \cell{1}{0.7}  & \cell{3}{2.7}  & \cell{15}{14.9} & \cell{0}{0.0}  & \cell{33}{33.2} & \cell{6}{5.8}  \\
			3.9 & Phantom riches & \cell{5}{4.5}  & \cell{28}{28.4} & \cell{12}{11.5} & \cell{43}{42.7} & \cell{2}{2.2}  & \cell{24}{23.7} & \cell{12}{12.2} \\
			3.10 & Comm.\ migration & \cell{2}{1.8}  & \cell{9}{9.3}  & \cell{3}{3.4}  & \cell{23}{22.6} & \cell{47}{46.7} & \cell{13}{13.4} & \cell{30}{30.3} \\
			3.11 & Post-pay shift & \cell{4}{4.2}  & \cell{25}{24.8} & \cell{13}{12.9} & \cellb{54}{54.3} & \cell{0}{0.2}  & \cellb{63}{62.6} & \cell{30}{29.7} \\
			\bottomrule
		\end{tabular}
	\end{table*}
	
	\subsubsection{Evidence Thinness}
	\label{sec:evidence_thinness}
	
	We found stark variation in the forensic specificity of Y rationales across indicators (Table~\ref{tab:rq2_specificity}). We measured forensic specificity as the proportion of Y rationales that contained concrete, actionable identifiers (a specific platform name, a monetary amount, a date, or a named entity) as opposed to generic descriptions like ``the perpetrator impersonated police'' with no further detail. At the rich end, Communication migration (77.5\%) and Social proof (77.4\%) produced highly specific rationales because victims naturally named the platform they were directed to (``moved me from Tinder to WhatsApp'') or described concrete testimonial evidence. Consistency (64.2\%) and Reciprocity (49.8\%) were also relatively rich, as victims tended to report specific amounts for escalating payments or initial payouts.
	
	At the thin end, Pretext (11.4\%) stood out: of 1{,}218 cases where the LLM detected fabricated credentials or platforms, only 139 rationales contained a specific detail such as a website URL, a document type, or an agency name. The rest were generic. Fear (22.2\%) and Urgency (18.3\%) were similarly thin, as victims reported \emph{that} they were threatened or pressured but rarely described \emph{how} (e.g., the exact wording, the claimed legal basis, or the stated deadline).
	
	This variation constituted a \emph{forensic detail gap}: the LLM reliably detected which tactics were used ($\bar{\kappa} = 0.69$), but for several indicators the victim narrative provided too little detail for investigators to act on the detection. The gap was not uniform. Communication migration and Social proof were already forensically rich, while Pretext, Fear, and Urgency would require elicitation to become investigatively useful.
	
	\begin{table}[t]
		\centering
		\caption{Forensic specificity of Category~3 Y rationales: proportion containing actionable details (platform, amount, date, or named entity).}
		\label{tab:rq2_specificity}
		\footnotesize
		\begin{tabular}{@{}llrc@{}}
			\toprule
			\textbf{Q\#} & \textbf{Indicator} & \textbf{Y count} & \textbf{Specific (\%)} \\
			\midrule
			3.10 & Comm.\ migration & 1{,}211 & 77.5 \\
			3.8 & Social proof & 1{,}220 & 77.4 \\
			3.7 & Consistency & 3{,}068 & 64.2 \\
			3.6 & Reciprocity & 1{,}432 & 49.8 \\
			3.11 & Post-pay shift & 3{,}290 & 41.6 \\
			3.4 & Liking & 964 & 39.4 \\
			3.9 & Phantom riches & 2{,}050 & 30.9 \\
			3.1 & Authority & 3{,}698 & 29.4 \\
			3.2 & Fear & 1{,}426 & 22.2 \\
			3.3 & Urgency & 1{,}071 & 18.3 \\
			3.5 & Pretext & 1{,}218 & 11.4 \\
			\bottomrule
		\end{tabular}
	\end{table}
	
	\subsubsection{Indicator Information Ceiling}
	\label{sec:info_ceiling}
	
	The previous section addressed indicators that are frequently \emph{detected} but produce thin evidence. A different bottleneck affected indicators that were rarely detected in the first place, not because the tactic was absent, but because victims did not mention it unprompted. We call this an \emph{information ceiling}.
	
	We observed this ceiling most clearly for Social proof (Q3.8). Cryptocurrency Investment Fraud routinely employs group-chat earnings screenshots and fabricated testimonials~\cite{franceschini2023Compound}, yet only 14.9\% of Crypto Investment Fraud cases received a Y label for Social proof. The likely explanation is that victims did not recognize these displays as a manipulation tactic worth reporting. The information was in the victim's memory, but not in the narrative.
	
	Liking (Q3.4) exhibited a related ceiling. Even in Romance Scams, where emotional bonding is the defining feature, only 4.9\% of cases produced a Y annotation. Victims may not have described rapport-building in terms that qualified as explicit perpetrator actions under the evidence-only annotation standard, or they may have considered emotional details too personal to report.
	
	\subsubsection{Cryptocurrency Evidence Gaps (Category~4)}
	
	Of the 10{,}994 cases, 10.7\% ($N_{\text{crypto}} = 1{,}172$) involved cryptocurrency. We found that cryptocurrency involvement was itself a strong differentiator ($V = 0.790$): 96.2\% of Cryptocurrency Investment Fraud cases involved crypto, versus 10.6\% of Romance Scams and 4.0\% of Sextortion.
	
	However, even when cryptocurrency was involved, the evidence needed for blockchain tracing was largely absent. Across all crypto-involved cases, only 36.9\% identified the cryptocurrency type, 21.5\% provided a recipient wallet address, and just 6.2\% included a transaction hash (Table~\ref{tab:rq2_crypto_coverage}). The gap was most acute in Cryptocurrency Investment Fraud ($n = 777$, 66.3\% of the crypto subset), which had the \emph{lowest} evidence availability despite being the fraud type most dependent on cryptocurrency. Only 4.0\% of these cases included a transaction hash and only 13.4\% provided a wallet address. Without these on-chain identifiers, blockchain tracing techniques such as wallet clustering and cross-complaint address linking~\cite{frowis2020Safeguarding} could not be applied to these cases.
	
	\begin{table}[t]
		\centering
		\caption{Cryptocurrency evidence: Y-rates and coverage among crypto-involved cases ($N_{\text{crypto}} = 1{,}172$).}
		\label{tab:rq2_crypto_coverage}
		\footnotesize
		\begin{tabular}{@{}clcc@{}}
			\toprule
			\textbf{Q\#} & \textbf{Evidence element} & \textbf{Y (\%)} & \textbf{Cov.\ (\%)} \\
			\midrule
			4.2 & Cryptocurrency type & 36.9 & 54.0 \\
			4.4 & Recipient wallet addr. & 21.5 & 52.6 \\
			4.3 & Transaction hash & 6.2 & 48.8 \\
			4.5 & Crypto ATM/kiosk used & 1.6 & 99.9 \\
			\bottomrule
		\end{tabular}
	\end{table}
	
	\section{Discussion}
    \label{sec:discussion}
	%% ============================================================
	
	Taken together, RQ1 and RQ2 painted a two-sided picture. On one side, RQ1 showed that different fraud types have distinct manipulation profiles, and that detection labels are sufficient to tell them apart. On the other, RQ2 revealed a \emph{forensic detail gap}: while the LLM can reliably detect which tactics were used, the victim narratives behind those detections vary widely in the actionable detail they provide. For some indicators, victims naturally report specifics that investigators can act on; for others, they give only vague descriptions. The following sections ask: what can binary profiles already achieve (Section~\ref{sec:triage}), how can we close the forensic detail gap (Section~\ref{sec:intake}), and what methodological challenges remain, including how the annotation methodology generalizes beyond fraud (Section~\ref{sec:characterization}).
	
	\subsection{How Far Can Differentiation Take Us?}
	\label{sec:triage}
    
	RQ1 showed that each major fraud type has a distinct manipulation profile. The statistical separation between these profiles is what we mean by \textit{differentiation}, which sets the ceiling on any downstream \textit{classification} task (i.e., assigning a case to a fraud type) or its operational instantiation at intake, \textit{triage} (i.e., routing complaints to a case linkage). Each case can be described by which of the 11 Category~3 tactics were present. Combined with the large effect sizes in Table~\ref{tab:rq1_fraud_type} and the fourfold range in tactic density (0.9 in Consumer Fraud to 3.6 in Cryptocurrency Investment Fraud), these profiles provide a strong basis for automated fraud-type triage.
	
	Beyond triage, these profiles have broader implications for fraud classification. They constitute a behavioral feature space that complements the technical and financial features traditionally used. Existing systems classify complaints by transaction metadata, contact channels, or keyword patterns. By contrast, the Category~3 profiles add an orthogonal dimension capturing \emph{how the perpetrator behaved}. As a concrete example, Cryptocurrency Investment Fraud and Business/Investment Fraud both showed high Consistency (58.5\% and 40.5\% of cases), but diverge sharply on Pretext (+82 percentage points), Post-payment behavioral shift ($-$8~pp), and Phantom riches (+19~pp). An automated system could use these divergences to flag probable pig-butchering cases among incoming complaints, even when transaction metadata alone is ambiguous.
	
	However, triage is only the first step. Once a complaint is routed to the right team, investigators need to act on it: seize the fabricated platform, trace the impersonated agency, link the case to others involving the same infrastructure. This requires the forensic detail that RQ2 showed is often missing. A Pretext = Y label tells the investigator that a fake platform was involved, but not \emph{which} platform. An Authority = Y label confirms impersonation, but not \emph{which} agency was impersonated. The next section addresses closing this gap.
	
	\subsection{Closing the Evidence Gap Through Intake Design}
	\label{sec:intake}
    
	The most direct way to close the forensic detail gap is to collect richer evidence at the point of reporting. The RQ2 rationale audit showed that Category~3 and Category~4 present complementary but distinct design challenges.
	
	For Category~3, the bottleneck is not coverage but \emph{forensic depth}. Indicators with high forensic specificity (Communication migration at 77.5\%, Social proof at 77.4\%, Consistency at 64.2\%) corresponded to concrete events that victims naturally described in detail. Indicators capturing fabricated artifacts (Pretext at 11.4\%) and temporal pressure (Urgency at 18.3\%) instead produced generic descriptions. Intake instruments could therefore include targeted follow-up prompts for the thin indicators: ``What evidence of legitimacy did they show you, such as screenshots, documents, or websites?'' (Q3.5) or ``What specific deadline or time pressure did they impose?'' (Q3.3).
	
	For indicators at the information ceiling (Section~\ref{sec:info_ceiling}), the challenge is more fundamental. Victims do not spontaneously report social proof (Q3.8) because they may not recognize group-chat earnings displays as a manipulation tactic. Intake forms could therefore include explicit prompts such as ``Did anyone else appear to be profiting from the same opportunity?'' to elicit information that victims possess but do not volunteer.
	
	While the Category~3 gaps could be narrowed through better prompting, Category~4 presents a more fundamental challenge. Blockchain forensic techniques such as wallet clustering and cross-complaint address linking~\cite{frowis2020Safeguarding} assume that on-chain identifiers are available as input. Our RQ2 findings showed that this assumption rarely held in practice: even in the fraud type most dependent on cryptocurrency, only 4\% of cases included a transaction hash and 13\% provided a wallet address. Whether this gap reflects victim unfamiliarity with blockchain details, platform obfuscation, or simply that intake instruments do not prompt for this information remains an open question. Targeted prompts such as ``Do you have the recipient's wallet address or the transaction confirmation from your exchange?'' could narrow these gaps if the primary barrier is elicitation rather than availability.
	
	Beyond static forms, the hybrid annotation methodology points to AI-assisted adaptive intake: an LLM-powered reporting interface that generates follow-up questions while the victim narrates, using the RQ1 prevalence profiles (Table~\ref{tab:prevalence}) as conditional logic. For instance, Consistency and Phantom riches co-occur in 30.9\% of Cryptocurrency Investment Fraud cases, far more than in any other fraud type. Detecting this pair mid-report could therefore trigger cryptocurrency-specific prompts for wallet addresses and transaction hashes, targeting the evidence gap most acute in that fraud type.
    
	\subsection{From Detection to Characterization}
	\label{sec:characterization}
    
	Even with improved intake, a methodological challenge remains: the current annotation pipeline determined whether a tactic was used (i.e., detection) but not what specific form it took (i.e., characterization). The substantial LLM-human agreement ($\bar{\kappa} = 0.69$, Table~\ref{tab:human_eval}) confirmed that the LLM could reliably answer ``was this tactic present?'' The evidence thinness findings, however, showed that forensic value ultimately depends on a harder question: ``what exactly did the perpetrator do?''
	
	This gap between detection and characterization is not uniform across indicators. Urgency (Q3.3) illustrated the hard end: it showed weak LLM-human agreement ($\kappa = 0.59$), among the weakest fraud-type associations ($V = 0.263$), low prevalence (maximum 20.9\%), and low forensic specificity (18.3\%). This convergence across all metrics suggested an indicator whose signal is inherently diffuse in victim narratives rather than a model-specific limitation. By contrast, Communication migration (77.5\% specificity) and Consistency (64.2\%) yielded characterization almost as a byproduct of detection, because victims naturally named the platforms and amounts involved. Future annotation pipelines could exploit this variation by applying lightweight extraction for high-specificity indicators and reserving more intensive methods, or intake-side elicitation, for thin indicators such as Pretext and Urgency. More broadly, the tiered annotation strategy is not specific to fraud: the same approach could be applied to other domains where investigators extract structured indicators from unstructured narratives, such as threat intelligence reports, vulnerability disclosures, or abuse complaints, with the indicator-level $\kappa$ protocol serving as a validation template.
	
	\subsection{Limitations and Future Work}
	
	Our research has several limitations. Our corpus of publicly available reports may not represent all victimization, and the schema captures only what victims report, not perpetrator-side artifacts. RQ1 associations are cross-sectional, not causal. Because public sources differ in intake forms, jurisdiction, and fraud-type composition, observed manipulation profiles may partly reflect reporting practices rather than perpetrator behavior alone. The multi-source design reduces dependence on a single venue, but does not eliminate this bias.
	
	On the annotation side, LLM annotation may miss subtle indicators, particularly implicit emotional coercion. The low Liking Y-rate (4.9\% even in Romance Scams, Section~\ref{sec:info_ceiling}) likely reflects the evidence-only annotation standard and victims' reporting tendencies, though LLM insensitivity to implicit cues may also contribute. The annotation results are validated for one LLM configuration and a stratified human audit subset. Future work needs to test their robustness across LLM families and estimate agreement with narrower confidence intervals.
	
	Finally, perpetrators aware of forensic profiling might vary their behavioral signatures, though the schema's breadth across 11 indicators raises the cost of evasion. Future work could validate the schema in operational law enforcement settings and develop structured intake instruments informed by the evidence audit findings.
	
	%% ============================================================
	\section{Conclusion}
	%% ============================================================
	
	We presented a forensic schema of 4 categories and 35 questions bridging persuasion research and forensic investigation of cyber fraud. Applied to 10{,}994 victim reports via LLM-driven annotation, our analysis yielded two findings. Different fraud types exhibited distinct psychological manipulation profiles (RQ1), but victim narratives varied widely in the actionable detail behind those detections (RQ2). This forensic detail gap was most acute for blockchain evidence, where on-chain identifiers were nearly absent. AI-assisted victim intake, where schema-informed follow-up questions dynamically elicit the missing detail, is the most direct path from detection to investigative impact. The tiered annotation strategy and validation protocol demonstrated here provide a reusable template for LLM-based structured extraction from other forensic text domains.

    \section*{Acknowledgment}
    The authors thank the anonymous reviewers for their constructive feedback. 
    This work was partially supported by the National Science Foundation (NSF) with award numbers 1921576, 2334196, and 2334197.
    
	%% ============================================================
	\bibliographystyle{IEEEtran}
	\bibliography{references}
	
	%% ============================================================
	\appendix
	
	\subsection{Annotation Prompt Template}
	\label{app:prompt}
	
	The Tiers~2--4 questions use Claude Haiku~4.5 API for LLM annotation. We present the prompts below in condensed form.
	
	\smallskip\begingroup\raggedright\scriptsize\ttfamily\noindent \textbf{a) Tier~1 deterministic:} 14 questions are resolved by a Python script before any LLM call. If the corresponding structured field is non-empty the answer is Y with the field value; otherwise N. The mappings are:\par\noindent
	Q1.1 $\rightarrow$ contact\_platform;\quad Q1.2 $\rightarrow$ contact\_method;\quad Q1.3 $\rightarrow$ scammer\_name;\quad Q1.4 $\rightarrow$ scammer\_phone;\quad Q1.5 $\rightarrow$ scammer\_email;\quad Q1.7 $\rightarrow$ scammer\_website;\quad Q1.12 $\rightarrow$ transaction\_date;\quad Q1.13 $\rightarrow$ transaction\_amount;\quad Q1.15 $\rightarrow$ payment\_method;\quad Q2.2 $\rightarrow$ bank\_name;\quad Q4.1 $\rightarrow$ crypto\_involved;\quad Q4.2 $\rightarrow$ crypto\_type;\quad Q4.3 $\rightarrow$ tx\_hash;\quad Q4.4 $\rightarrow$ recipient\_wallet\par
	\endgroup
	
	\smallskip\begingroup\raggedright\scriptsize\ttfamily\noindent \textbf{b) Tier~2/3 prompt:} You are a forensic analyst annotating cyber fraud victim reports. Each case has structured fields and a narrative description. Annotate 10 questions across Tier~2 and~3 using BOTH sources.\par
	\smallskip\noindent
	\textit{Tier 2 Field + narrative (4 questions):} Check the structured field first. If non-empty, answer Y with the field value. If empty, search the narrative.\par\noindent
	Q1.6 Other contact identifiers (sp/se + narrative); Q1.14 Total loss (ta + narrative); Q2.1 Traditional banking (bn/ba + narrative); Q2.3 Recipient identification (ba + narrative)\par
	\smallskip\noindent
	\textit{Tier 3 Narrative extraction (6 questions):} No structured field. Extract from narrative.\par\noindent
	Q1.8 Subject address; Q1.9 Subject IP; Q1.10 Multiple perpetrators; Q1.11 Coordinated group operations; Q1.16 Routing instructions; Q4.5 Crypto ATM (N if Q4.1=N)\par
	\smallskip\noindent
	For each question answer Y/N/U with a JSON object. Rationale required for Y answers, null for N/U.\par
	\endgroup
	
	\smallskip\begingroup\raggedright\scriptsize\ttfamily\noindent \textbf{c) Tier~4 prompt:}
	You are annotating fraud case descriptions for Q3: Persuasion \& Manipulation Techniques. Evaluate 11 indicators (Q3.1--Q3.11). This is an evidence-only task: label Y only when the narrative contains explicit perpetrator actions or utterances; provide an exact supporting quote (15--60 words) for every Y. No inference, no category-based reasoning.\par
	\smallskip\noindent
	\textit{Input:} CSV with columns fraud\_category, case\_id, date, description.\par\noindent
	\textit{Output:} JSON array; each element: \{"case\_id": "...", "Q3.X": ["Y"|"N"|"U", "quote or null"]\}. U reserved for genuine ambiguity ($<$5\%).\par
	\smallskip\noindent
	\textit{Indicator definitions:}\par\noindent
	Q3.1 Authority --- claims a professional role or organizational affiliation.\par\noindent
	Q3.2 Fear --- explicit threats of serious consequences (legal, physical, exposure).\par\noindent
	Q3.3 Urgency --- explicit deadlines or immediate demands.\par\noindent
	Q3.4 Liking --- builds romantic interest, friendship, or emotional bond.\par\noindent
	Q3.5 Pretext --- impersonates a real entity or creates fabricated artifacts (fake documents, spoofed emails, scam websites). Excludes investment/trading apps.\par\noindent
	Q3.6 Reciprocity --- personally gives something (gift, service, intimate content, prize) to create obligation. Excludes investment platform returns.\par\noindent
	Q3.7 Consistency --- after initial payment, demands additional money via new pretextual fees or charges.\par\noindent
	Q3.8 Social Proof --- demonstrates others' success via visible evidence (screenshots, testimonials).\par\noindent
	Q3.9 Phantom Riches --- specific promise of unusually high or guaranteed returns.\par\noindent
	Q3.10 Communication Migration --- directs victim to a different communication channel or scammer-managed group. Excludes investment/trading apps.\par\noindent
	Q3.11 Post-Payment Shift --- after payment, scammer becomes unreachable or platform goes offline.\par
	\smallskip\noindent
	\textit{Key disambiguation rules:}\par\noindent
	Account freeze for fees = Q3.7; freeze tied to legal accusation = Q3.2.\par\noindent
	Still reachable demanding more = Q3.7; vanished entirely = Q3.11; both can co-occur.\par\noindent
	``Unable to withdraw'' alone = Q3.7, not Q3.11; ``realized fraud'' alone $\neq$ Q3.11.\par
	\endgroup
	
	\subsection{Categories 1--2 Question Listing}
	\label{app:cat12}
	
	Table~\ref{tab:cat12_questions} lists all Category~1 and~2 questions.
	
	\begin{table}[ht]
		\centering
		\caption{Categories 1--2: Forensic Metadata and Financial Flow.}
		\label{tab:cat12_questions}
		\scriptsize
		\begin{tabular}{@{}lp{6.2cm}@{}}
			\toprule
			\textbf{Q\#} & \textbf{Operationalized Question} \\
			\midrule
			\multicolumn{2}{@{}l}{\textit{Category~1: Standard Forensic Metadata}} \\
			1.1 & Did it include the contact platform used by the perpetrator? \\
			1.2 & Did it describe how the perpetrator initially contacted the victim? \\
			1.3 & Did it include the perpetrator's name or entity? \\
			1.4 & Did it include the perpetrator's telephone number? \\
			1.5 & Did it include the perpetrator's email address? \\
			1.6 & Did it include other contact identifiers for the perpetrator? \\
			1.7 & Did it mention a web domain or URL associated with the fraud? \\
			1.8 & Did it include the perpetrator's physical address? \\
			1.9 & Did it include the perpetrator's IP address? \\
			1.10 & Were multiple perpetrators mentioned (e.g., organized group, named accomplices)? \\
			1.11 & Were multiple coordinated group operations mentioned (e.g., distinct roles, coordination structure, or division of labor)? \\
			1.12 & Was the transaction date mentioned? \\
			1.13 & Was the transaction amount mentioned? \\
			1.14 & Was the total financial loss quantified? \\
			1.15 & Was the payment method mentioned (e.g., bank transfer, credit card, cryptocurrency, gift card, wire transfer)? \\
			1.16 & Did the perpetrator provide routing or transfer instructions to the victim? \\
			\midrule
			\multicolumn{2}{@{}l}{\textit{Category~2: Traditional Financial Flow}} \\
			2.1 & Did the transaction involve traditional banking methods (e.g., wire transfer, bank account, check)? \\
			2.2 & Was the name of the receiving bank identified? \\
			2.3 & Was the recipient identified by bank account name or entity? \\
			\bottomrule
		\end{tabular}
	\end{table}
	
\end{document}